\RequirePackage{fix-cm}
\documentclass[smallextended]{svjour3}       
\smartqed  
\usepackage{svg}
\usepackage[super]{cite} 
\usepackage{epstopdf}
\usepackage{caption}
\usepackage{setspace}
\usepackage{amssymb}
\usepackage{multirow}
\usepackage{booktabs}
\usepackage{amsmath}
\usepackage{amssymb}
\usepackage{mathrsfs}
\usepackage{longtable}
\usepackage{lscape}
\usepackage{hyperref}
\graphicspath{{figs/}{figsgaoerb/}} 
\usepackage{url}
\usepackage{tabularx}
\usepackage{epsfig}
\usepackage{colortbl}
\usepackage{subfigure}
\usepackage{tikz,mathpazo}
\usepackage{graphicx}
\usepackage[noend]{algpseudocode}
\usepackage{algorithm}
\usepackage{algorithmicx}   
\usepackage{array}  
\usepackage{hyperref}
\usepackage[misc]{ifsym}
\begin{document}
	
	\title{Quantum soft likelihood function based on ordered weighted average operator  }
	
	
	\author{Tianxiang Zhan        \and
		Yuanpeng He               \and
		Fuyuan Xiao\textsuperscript{*} 
	}
	
	
	\institute{Tianxiang Zhan \at
		School of Computer and Information Science Southwest University Chongqing, China\\
		School of Big Data and Software Engineering, Chongqing University, Chongqing, China\\
		\email{zhantianxiangswu@163.com}           
		\and
		Yuanpeng He \at
		School of Computer and Information Science Southwest University Chongqing, China\\
		School of Big Data and Software Engineering, Chongqing University, Chongqing, China\\
		\email{heyuanpengswu@163.com}           
		\and
		\Letter Fuyuan Xiao\at
		School of Big Data and Software Engineering, Chongqing University, Chongqing, China\\
		\email{xiaofuyuan@cqu.edu.cn; doctorxiaofy@hotmail.com} 
	}
	
	\date{Received: date / Accepted: date}

\maketitle

\begin{abstract}
		Quantum theory is the focus of current research. Likelihood functions are widely used in many fields. Because the classic likelihood functions are too strict for extreme data in practical applications, Yager proposed soft ordered weighted average (OWA) operator. In the quantum method, probability is represented by Euler's function. How to establish a connection between quantum theory and OWA is also an open question. This article proposes OWA opreator under quantum theory, and discusses the relationship between quantum soft OWA operater and classical soft OWA operator through some examples. Similar to other quantum models, this research has more extensive applications in quantum information. 
\end{abstract}

\section{Introduction}
In functional applications such as fuzzy sets \cite{zimmermann2000application,fei2019new,xiao2019distance}, rough sets \cite{dubois1990rough,qiao2021io,bashir2021conflict}, Z‐numbers \cite{zadeh2011note,kang2018method,tao2020gmcdm,song2020fr}, belief structures \cite{yager2018fuzzy,han2019novel}, D numbers \cite{xiao2019multiple,zhao2019performer,xiao2018novel}, entropy function \cite{deng2016deng,cui2019improved,wang2017rumor}, and evidence theory \cite{dempster2008upper,shafer1992dempster,https://doi.org/10.1002/int.22366}, the analysis of unknown facts is crucial. Probability has been thoroughly researched and widely applied in the area of uncertainty, including decision-making \cite{zhou2019evidential,sun2019new,zhou2018evidential,fei2019ds}, estimation \cite{meng2019enhanced,li2021multisource}, and uncertainty measurement \cite{abellan2017random,wang2018uncertainty,garg2016new,li2021improved}. The likelihood function is one of the most commonly used models in probability. When there is data available to describe the reasonableness of the parameter value, the likelihood function can be used. However, it is very strict in practical applications, because the original likelihood function composed of probability products is too strict to show the compatible probability of different evidences, because any low probability will greatly reduce the likelihood function. A typical soft likelihood function proposed by Yager is based on ordered weighted average (OWA) to solve this problem. OWA can avoid the problem of low-value lowering the likelihood function, and has played a huge role in the processing of uncertain information. 

In the soft likelihood function, there are many operations involving the superposition of different probabilities. Through quantum theory, it is a good perspective to study the superposition effect. Quantum mechanics can consider the connection between different events. There are non-classical correlations between the physical properties of quantum systems, so that these quantities cannot be accurately obtained, but they exist as linear superposition state vectors. Based on the nature of quantum systems, quantum information has gradually developed and been used in many fields, such as reliability analysis, intelligent system, decision making and so on.

Quantum theory is a topic worthy of discussion, and it has applications in many fields such as cybersecurity \cite{mosca2018cybersecurity,denning2019quantum,abellan2018future}, drug development \cite{lam2020applications,zhou2010quantum,cao2018potential}, financial modeling \cite{martin2021toward,castelluccio2021quantum,dong2021infinite}, better  batteries \cite{kou2021hybridizing,jiao2021mo2c}, traffic optimization \cite{zhang2021new,grant2021benchmarking,guo2021capacity}, weather forecasting \cite{wang2021intelligent} and so on\cite{lai2020parrondo, Dai2020interferenceQLBN,gao2021QPFET}. Quantum probability theory has many similarities with the assumptions and methods of classical probability theory. In a sense, quantum probability is an extension of classical probability. Therefore, applying OWA to quantum mechanics can better analyze the superposition effect and strengthen the application of the likelihood function under quantum probability. This paper proposes OWA under quantum theory, which expands the application of soft likelihood function in quantum theory.

The structure of the paper is as follows. In the section 2, a brief introduction to the basic knowledge of OWA and Quantum Mass Function. The section 3 proposes Quantum soft likelihood function based on ordered weighted average operator. In the section 4, some numerical examples illustrate the calculation details. Section 5 summarizes the quantum soft likelihood function based on ordered weighted average operator.
\section{Preliminaries}
\subsection{Original likelihood function}
The definition of the likelihood function is the product of probabilities, calculated as follows:
$$Prod\left(i\right)=\prod_{k=1}^{i}p_k \eqno(1)$$
where $Prod\left(i\right)$ is likelihood function for probabilities $p_k$ of events $A_i$.

\subsection{Quantum mass function}
The quantum recognition framework and quantum mass function \cite{gao2020quantum} are defined as follows:

A frame of discernment indicates as X which is a set of mutually exclusive and collectively exhaustive events $A_i$:
$$X=\left\{A_1,A_2,A_3,...,A_n\right\} \eqno(2)$$
where the quantity of the events is $n$. The power set of $X$ is indicated as:
$$2^X= \left\{\emptyset,\left\{A_1\right\},\left\{A_1,A_2\right\},...,X\right\} \eqno(3)$$
where $2^X$ composed of $2^n$ elements.

In quantum frame of discernment, the quantum mass function $m$ is defined as follows:
$$m\left(A\right)=ae^{\theta i} \eqno(4)$$
where the mass function satisfies:
$$|m\left(A\right)|=a^2 \eqno(5)$$
$$|m\left(\emptyset\right)|=0  \eqno(6)$$
$$\sum_{A\subseteq X}|m\left(A\right)|=1  \eqno(7)$$

\subsection{Ordered weighted average operator}
The OWA operator was first proposed by Yager to create a likelihood function \cite{yager1993families,yager2012ordered} , and then combine multiple probabilities to play a role in data fusion.
The OWA operator is an n-dimensional vector to one-dimensional mapping, which is defined as follows:
$$OWA\left(A_1,A_2,...,A_n\right)=\sum_{j=1}^{n}\omega_jB_j \eqno(8)$$
$$\sum_{j=1}^{n}\omega_j=1 \eqno(9)$$
where $B_j$ is $j-th$ largest elements in $\left\{A_1,A_2,...,A_n\right\}$. The two characterization methods related to OWA operators are attitude characteristics and dispersion, so the weighted variable $\omega_j$ could be defined as follows:
$$\omega_j=\left(\frac{i}{n}\right)^{\left(1-\alpha\right)/\alpha}-\left(\frac{i-1}{n}\right)^{\left(1-\alpha\right)/\alpha} \eqno(10)$$
where $\alpha$ is the degree of optimism expected. The smaller the $\alpha$, the more pessimistic, $\alpha=0.5$ represents a neutral position.

\section{Quantum ordered weighted average operator}
In this section, a new likelihood function is proposed based on the OWA operator in quantum theory. Compared with the traditional OWA operator, the quantum OWA operator takes into account the superposition effect between events. Classic OWA is a special case of quantum OWA.
\subsection{Limitations of existing OWA operands}
The classic probability does not take into account the angle between the two probabilities. For calculations, the classic probability is a scalar, as shown in the Fig.1 below:

\begin{figure}[htbp]
	\centerline{\includegraphics[scale=1.2]{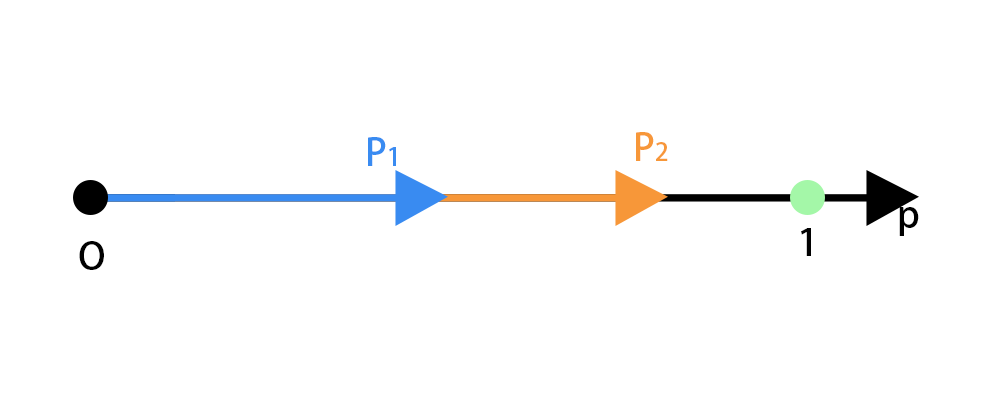}}
	\caption{The classical probability $P_1$(blue vector),$P_2$(yellow vector) has no angle}
\end{figure}
The classical probability can be regarded as a special vector with a constant angle of 0. Compared with quantum probability, classical probability does not have the ability to express an angle, and quantum probability can express an angle, as shown in the Fig.2:

\begin{figure}[htbp]
	\centerline{\includegraphics[scale=0.2]{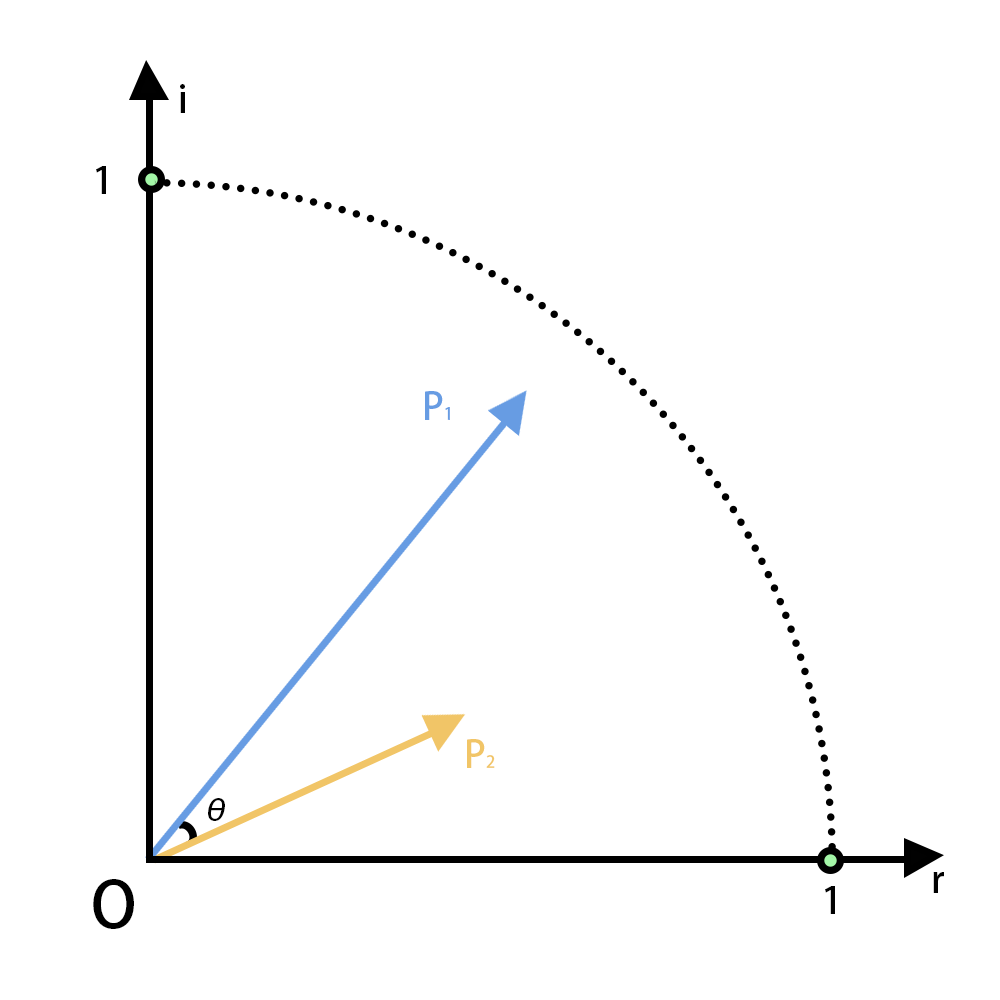}}
	\caption{There is an angle between quantum probability $P_1$ and $P_2$}
\end{figure}
In this case, the classical OWA operator is difficult to deal with quantum probability.

\subsection{Quantum ordered weighted average operator}
The quantum OWA operator defines the likelihood function of a new quantum theory. The classic OWA operator needs to sort the existing probabilities according to the magnitude of the probability, and assign different weights according to the order of magnitude. The quantum probability is not a scalar, and in most cases, it is impossible to directly compare the magnitude of the value. In the quantum probability comparison, the method of comparing the module length is adopted, which is consistent with the comparison method of classical probability. The quantum OWA operator is defined as follows.

A set of mutually exclusive and collectively exhaustive events indicates as follows:
$$X=\left\{A_1,A_2,...,A_n\right\} \eqno(11)$$
The quantum probability of $A_i$ is expressed as:
$$p_q\left(A\right)=ae^{\theta i} \eqno(12)$$
where $\theta$ is the angle. According to Euler formula, the quantum probability $p\left(A\right)$ can be written in the following form:
$$p_q\left(A\right)=ae^{\theta i}=a\left(cos\left(\theta\right)+i\ sin\left(\theta\right)\right) \eqno(13)$$
Then the modulus length of quantum probability is calculated as follows:
$$|p_q\left(A\right)|=\sqrt{a^2\bullet\left(cos^2\left(\theta\right)+sin^2\left(\theta\right)\right)}=\sqrt{a^2}=a \eqno(14)$$
The set $X$ arranged in descending order according to the quantum probability modulus length is expressed as:
$$X=\left\{A_1',A_2',...,A_n'\right\} \eqno(15)$$
where $A_i'$ is the $i-th$ largest events in the set $X$. So the quantum OWA operator is calculated as follows:
$$QOWA\left(A_1,A_2,...,A_n\right)=\sum_{i=1}^{n}\omega_iA_i'=\sum_{i=1}^{n}\left(\left(\frac{i}{n}\right)^{\frac{1-\alpha}{\alpha}}-\left(\frac{i-1}{n}\right)^{\frac{1-\alpha}{\alpha}}\right)\bullet a_i'e^{\theta'\ i} \eqno(16)$$

\subsection{Compatibility of Quantum ordered weighted average operator}
When the quantum probability of all events of the quantum OWA operator is 0, the quantum probability of event $A$ is calculated as follows:
$$p_q\left(A\right)=a\left(cos\left(o\right)+i\ sin\left(o\right)\right)=a \eqno(17)$$
$$|p_q\left(A\right)|=\sqrt{a^2\bullet\left(cos^2\left(0\right)+sin^2\left(0\right)\right)}=\sqrt{a^2}=a \eqno(18)$$
where the quantum ordered weighted average operator is calculated as follows:
$$QOWA\left(A_1,A_2,...,A_n\right)=\sum_{i=1}^{n}\omega_iA_i'=\sum_{i=1}^{n}\left(\left(\frac{i}{n}\right)^{\frac{1-\alpha}{\alpha}}-\left(\frac{i-1}{n}\right)^{\frac{1-\alpha}{\alpha}}\right)\bullet a_i' \eqno(19)$$

At this time, since the angles of all quantum probabilities are 0, the quantum probability can be regarded as a classical probability, so the OWA operator is calculated as follows:
$$OWA\left(A_1,A_2,...,A_n\right)=\sum_{i=1}^{n}\omega_iA_i'=\sum_{i=1}^{n}\left(\left(\frac{i}{n}\right)^{\frac{1-\alpha}{\alpha}}-\left(\frac{i-1}{n}\right)^{\frac{1-\alpha}{\alpha}}\right)\bullet a_i' \eqno(20)$$
$$OWA\left(A_1,A_2,...,A_n\right)=QOWA\left(A_1,A_2,...,A_n\right) \eqno(21)$$
It can be seen that the results of the quantum OWA operator and the classical OWA operator are consistent. Quantum QOWA is an extended form of OWA. When the quantum probability angle of all events is 0, QOWA degenerates into classic OWA.

The algorithm shows steps to obtain quantum soft likelihood function based on ordered weighted average operator as follows:

\begin{algorithm}[htbp]
	\caption{Quantum soft likelihood function based on ordered weighted average operator}
	\label{alg::conjugateGradient}
	\begin{algorithmic}[1]
		\Require
		A frame of discernment $X$
		\Ensure
		Soft likelihood function P
		\State Compute the products of quantun probabilities
		\State Compute the weights of each events
		\State Combine the weighted probability
	\end{algorithmic}
\end{algorithm}

\section{The computation of quantum soft likelihood function based on ordered weighted average operator}
There is an assumption that there are 5 $\left(n=5\right))$ sources of evidence. It supposed that suspect A whose probability with the seven sources of evidence is:
$$E=\left\{p_1=0.3-0.7i,p_2=0.4-0.9i,p_3=0.5+0.3i,p_4=0.6+0.8i,p_5=0.2+0.5i\right\} \eqno(22)$$
$$|p|<=1 \eqno(23)$$

The modulus length of each quantum probability is easy to calculate and show in the Tab.1:
\begin{table}[htbp]
	\setlength{\tabcolsep}{20mm}
	\centering
	\begin{tabular}{cc}
		\hline
		Probability & Modulus length \\ \hline
		$p_1$       & 0.7616         \\
		$p_2$       & 0.9849         \\ 
		$p_3$       & 0.5831         \\ 
		$p_4$       & 1              \\ 
		$p_5$       & 0.5385         \\ \hline
	\end{tabular}
	\caption{The modulus length of each quantum probability}
\end{table}

According to the modulus length of each probability, the order of each probability is shown in Tab.2:
\begin{table}[htbp]
	\setlength{\tabcolsep}{20mm}
	\centering
	\begin{tabular}{cc}
		\hline
		Order & Probability \\ \hline
		1     & $p_4$       \\ 
		2     & $p_2$       \\ 
		3     & $p_1$       \\ 
		4     & $p_3$       \\ 
		5     & $p_5$       \\ \hline
	\end{tabular}
	\caption{The order of each quantum probability}
\end{table}

\subsection{Step 1: Compute the product of quantun probabilities}
Since the product of all probabilities is the original form of the likelihood function, it's important to measure the split product of ordered probabilities first, then merge them with the aggregation operator to soften the likelihood function to get the probability product seen as follows:

$$Prod(1)=(0.6+0.8i)=0.6+0.8i \eqno(24)$$
$$Prod(2)=(0.6+0.8i)(0.4-0.9i)=0.96-0.22i \eqno(25)$$
$$Prod(3)=(0.6+0.8i)(0.4-0.9i)(0.3-0.7i)=0.134-0.738i \eqno(26)$$
$$Prod(4)=(0.6+0.8i)(0.4-0.9i)(0.3-0.7i)(0.5+0.3i)=0.2884-0.3288i \eqno(27)$$
$$Prod(5)=(0.6+0.8i)(0.4-0.9i)(0.3-0.7i)(0.5+0.3i)(0.2+0.5i)=0.2221+0.0784i \eqno(28)$$

\subsection{Step 2: Compute the weights of each events}
According to the definition of $\omega_i$ in Section 2, for different degrees of optimism, each group of $\omega_i$ values is different. Here take $\alpha=0.2$ (relatively pessimistic), $\alpha=0.5$ (neutral) and $\alpha=0.8$ (relatively optimistic) as examples to calculate $\omega_i$, and the results are shown in Tab.3:
\begin{table}[htbp]
	\setlength{\tabcolsep}{10mm}
	\centering
	\begin{tabular}{cccc}
		\hline
		& $\alpha=0.2$ & $\alpha=0.5$ & $\alpha=0.8$ \\ \hline
		$\omega_1$ & 0.0016       & 0.2          & 0.6687       \\ 
		$\omega_2$ & 0.024        & 0.2          & 0.1265       \\ 
		$\omega_3$ & 0.104        & 0.2          & 0.0848       \\ 
		$\omega_4$ & 0.28         & 0.2          & 0.0656       \\ 
		$\omega_5$ & 0.5904       & 0.2          & 0.0543       \\ \hline
	\end{tabular}
	\caption{The order of each initial likelihood function values}
\end{table}

\subsection{Step 3: Combine the weighted probability}
The weighted calculation of the weight and the original likelihood function is the proposed quantum soft likelihood function. For three different degrees of optimism, the proposed quantum soft likelihood function values are shown in Tab.4:

\begin{table}[htbp]
	\setlength{\tabcolsep}{5mm}
	\centering
	\begin{tabular}{cccc}
		\hline
		$\alpha$   & $\alpha=0.2$   & $\alpha=0.5$ & $\alpha=0.8$   \\ \hline
		Likelihood & 0.2498-0.1266i & 0.4409-0.0817i       & 0.565+0.4273i \\ \hline
	\end{tabular}
	\caption{Likelihood function values in example optimistic attitudinal character}
\end{table}

For different degrees of optimism, calculate the likelihood function of different degrees of optimism with 0.05 as the step size, see Tab.5:

\begin{table}[htbp]
	\centering
	\begin{tabular}{cc}
		\hline
		$\alpha$ & Likelihood     \\ \hline
		0.05 & 0.2231+0.0725i \\
		0.1  & 0.2297+0.0198i \\
		0.15 & 0.2369-0.0563i \\
		0.2  & 0.2498-0.1266i \\
		0.25 & 0.2728-0.1771i \\
		0.3  & 0.3036-0.2028i \\
		0.35 & 0.3387-0.2033i \\
		0.4  & 0.3745-0.181i  \\
		0.45 & 0.4089-0.1391i \\
		0.5  & 0.4409-0.0817i \\
		0.55 & 0.4695-0.0121i \\
		0.6  & 0.4948+0.0663i \\
		0.65 & 0.5166+0.1515i \\
		0.7  & 0.5355+0.2412i \\
		0.75 & 0.5515+0.3334i \\
		0.8  & 0.565+0.4273i  \\
		0.85 & 0.5765+0.5214i \\
		0.9  & 0.586+0.6154i  \\
		0.95 & 0.5937+0.7083i \\ \hline
	\end{tabular}
	\caption{Likelihood function values in different optimistic attitudinal character}
\end{table}

Likehood is displayed in three-dimensional coordinates as shown in Fig.3, where $(r,i)$ represents quantum probability, and $\alpha$ represents optimistic attitudinal character.
\begin{figure}[htbp]
	\centerline{\includegraphics[scale=0.4]{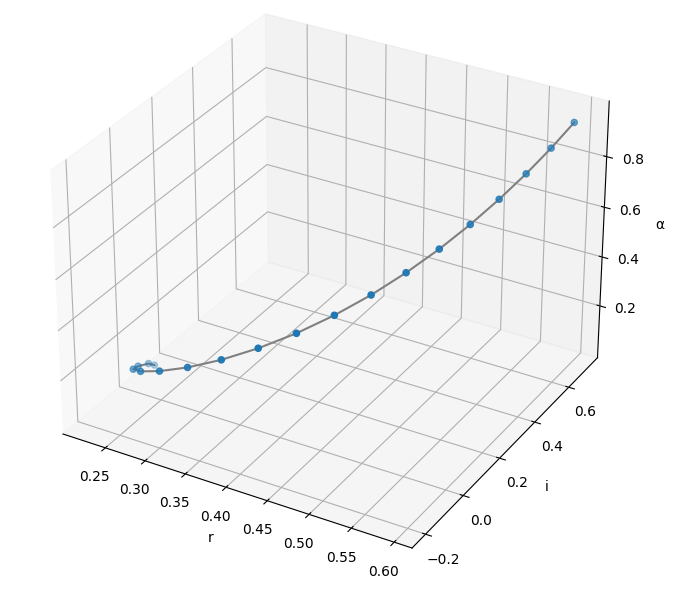}}
	\caption{Three-dimensional likelihood map}
\end{figure}

\subsection{Analysis}
The quantum OWA operator can be seen as an extension of the classic OWA operator. In the above example, the quantum OWA soft-likelihood function is compatible with the framework of the classic classic OWA soft-likelihood function. Quantum OWA is an extension of classic OWA, and classic OWA is a special case of quantum OWA. This article defines the calculation method of the soft likelihood function of OWA in the quantum domain for the first time, and expands the use of OWA operators.

\section{Conclusion}
A Quantum soft likelihood function based on ordered weighted average operator is proposed. The Quantum soft likelihood function based on ordered weighted average operator takes advantage of the OWA operator to complete the calculation of the quantum likelihood function, and considers the degree of support between the probabilities in the aggregation process. In short, the soft likelihood function proposed in this paper has broad application prospects in practical applications under uncertain environments.

\section*{Acknowledgment}
The authors greatly appreciate the reviewers' suggestions and the editor's encouragement. This research is supported by the National Natural Science Foundation of China (No.62003280).

\section*{Declarations}
\textbf{Conflict of interest} There is no conflict of interest.
\\
\textbf{Ethical approval} This article does not contain any studies with humanparticipants or animals performed by any of the authors.

\appendix
\bibliographystyle{spmpsci}
\bibliography{References}

\end{document}